\documentclass[preprint2]{aastex}

\shorttitle{TU Muscae}
\shortauthors{Terrell et al.}

\begin{document}


\title{Observational Studies of Early-type Overcontact Binaries: \\ TU Muscae}

\author{Dirk Terrell}
\affil{Department of Space Studies, Southwest Research Institute,
    \\1050 Walnut St., Suite 400, Boulder, CO 80302}
\email{terrell@boulder.swri.edu}

\author{Ulisse Munari}
\affil{Osservatorio Astronomico di Padova, Sede di Asiago, I-36032 Asiago (VI), Italy}
\email{munari@pd.astro.it}

\author{Toma\v{z} Zwitter}
\affil{Department of Physics, University of Ljubljana, Jadranska 19, 1000 Ljubljana, Slovenia}
\email{tomaz.zwitter@fmf.uni-lj.si}

\author{Robert H. Nelson}
\affil{1393 Garvin Street, Prince George, BC, Canada, V2M 3Z1}
\email{bob.nelson@shaw.ca}

\begin{abstract}
We present new spectroscopic and photometric data on the early-type overcontact
binary TU~Muscae. The analysis of the spectroscopic data shows that the line of
sight to the system crosses three kinematically sharp and well-separated interstellar
reddening sources and that the stars rotate synchronously. We present new radial
velocities that are in good agreement with earlier optical velocities and, thus, 
do not confirm the systematically smaller velocities obtained from IUE spectra.
The optical velocities are analyzed simultaneously with the photometric data to
derive accurate absolute dimesions for the binary components.The results show that 
TU~Mus consists of an O7.5 primary with $M_{1}=23.5 \pm 0.8 M_{\sun}$, 
$R_{1}=7.48 \pm 0.08 R_{\sun}$ and an O9.5 secondary with
$M_{2}=15.3 \pm 0.4 M_{\sun}$, $R_{2}=6.15 \pm 0.07 R_{\sun}$ in an overcontact 
configuration and that the orbital period has remained constant over the 
three decades covered by the observations. These results might imply that the mass 
transfer seen in late-type overcontact binaries does not occur in their 
early-type counterparts.
\end{abstract}


\section{Introduction}

The variability of \objectname{TU Muscae} (HD 100213) was discovered by \cite{oo28} and the system was 
first studied in detail by
\citet[hereafter AG75]{ag75} who list references to earlier work on this early-type
overcontact \citep{rew01} binary. Given the paucity of eclipsing, double-lined O-type binaries, 
TU Mus has received surprisingly little attention from observers. AG75 published fully covered
$uvby$ light curves and fourteen radial velocities for each star near quadratures from 
photographic spectra of moderate resolution. \citet[hereafter S95]{s95} presented twenty-four radial velocities for the primary star and twenty-three for the secondary from high resolution International Ultraviolet Explorer (IUE) spectra. 
TU~Mus was bright enough ($V_{T}=8.4$) to be observed by Hipparcos and a light curve with 
about a hundred points was published in the Hipparcos Epoch Photometry Annex \citep{hipp}. 
Two analyses of these data using modern light curve synthesis tools have been published: a 
preliminary report on the current work by \cite{dt02} and an analysis of the AG75 photometry 
by \cite{wr81}.

The two sets of radial velocities give quite discrepant results for the absolute dimensions. AG75 obtained photographic spectra with a dispersion of 20 \AA~mm$^{-1}$ and found 
$K_1$=251.3 $\pm$4.8 km~sec$^{-1}$ and $K_2$=370.5 $\pm$4.3 km~sec$^{-1}$. S95 found 
$K_1$=216.7 $\pm$2.7 km~sec$^{-1}$ and $K_2$=345.4 $\pm$3.1 km~sec$^{-1}$ from their IUE
spectra that had signal-to-noise ratios (S/N) of 10-20. Adopting an inclination of $76^{\circ}$, these numbers result 
in masses of $M_{1}=17.2 M_{\sun}$ and $M_{2}=10.8 M_{\sun}$ for the IUE velocities and
$M_{1}=23.5 M_{\sun}$ and $M_{2}=15.8 M_{\sun}$ for the optical velocities, a difference 
much larger than the error limits allow. In order to explore the nature of this 
discrepancy, we obtained high-resolution, high S/N spectra of TU~Mus near the quadratures.
Although we have a limited number of spectra due to the difficulty of obtaining observing time
on large telescopes for the study of eclipsing binaries, we find very good agreement with the
AG75 results. 

Given the good agreement between our velocities and those of AG75, we performed a simultaneous
solution of those velocities, the AG75 $uvby$ photometry, the Hipparcos photometry and 
new $BV$ photometry from the 2002 observing season. We used the 2003 version of the 
Wilson-Devinney (WD) program (\cite{wd71},\cite{rew79},\cite{rew90}) and present new elements for the 
binary, showing that the period has remained remarkably constant over the roughly thirty year 
span of all the observations. 

\section{Spectroscopy}

Nine ESO 2.2m + FEROS spectra have been
secured, in three groups of three spectra each (Table \ref{speclist}). The wavelength range of the
spectra extends from 3900~\AA\ to 9200~\AA\ with $R=48,000$ and a 400 second
exposure time. The S/N is evaluated 
around H$\beta$ and the orbital phase is computed with the ephemeris given in section 
\ref{analysis}.

\begin{deluxetable}{ccccc}
\tabletypesize{\scriptsize}
\tablecaption{ESO 2.2m + FEROS Spectra of TU Mus \label{speclist}}
\tablewidth{0pt}
\tablehead{
\colhead{Spectrum Number} & \colhead{Date} & \colhead{UT of mid-exposure}
           & \colhead{S/N}  & \colhead{Orbital Phase}
}
\startdata
1346, 1347, 1348 & Feb 22, 2003 & 07:08:56, 07:18:16, 07:27:31 &  75,  68,  72 & 0.032 \\   
1629, 1630, 1631 & Feb 25, 2003 & 06:07:24, 06:16:39, 06:25:59 & 160, 165, 158 & 0.211 \\
1693, 1694, 1695 & Feb 26, 2003 & 00:45:27, 00:54:47, 01:04:07 & 125, 130, 120 & 0.771 \\
\enddata
\end{deluxetable}

\subsection{Rotational velocity}

AG75 estimated the rotational velocities of the primary and secondary stars in TU~Mus 
as 285 and 240 km~sec$^{-1}$ ($\pm$10\%), while S95 derived
significantly slower rotations: 250 and 195 km~sec$^{-1}$ ($\pm$10\%),
respectively. To derive the rotational velocity we used the relation
\begin{equation}
\label{reddening}
V_{\rm rot} = 42.42 \times HIW\ -\ 35  {\rm km~sec^{-1}}
\end{equation}
\noindent
calibrated by \cite{mun99} on the width at half maximum of
\ion{He}{1}~5876~\AA\ in high resolution spectra of O and B stars. The decovolution
of the \ion{He}{1}~5876~\AA\ line profile in the two TU~Mus quadrature spectra gives a HIW of
7.26 and 7.35~\AA\ for the primary, and 6.37 and 6.36~\AA\ for the secondary,
which corresponds to 273 and 277, and 234 and 236~km~sec$^{-1}$
respectively. The Munari and Tomasella relation was calibrated using single, and thus axially 
symmetric, stars. The components of TU Mus depart quite strongly from axial symmetry so we 
investigated its influence on the HIW using the line profile feature of WD. We computed the
HIW for TU Mus and for a spherical star with radius equal to the $r_{back}$ radii of the TU Mus 
components. We found that the HIW for 
TU Mus was about 5\% smaller than for the spherical star. Applying this correction to our
data, we adopt 290 and 248~km~sec$^{-1}$ as the rotational velocities of primary and secondary. 
Of course, the concept of a rotational velocity is of less obvious usefulness for stars like
TU Mus than for axially symmetric ones. The question arises as to where on the equator of TU Mus
the above rotational velocities apply. Since our spectra were taken at quadrature phases, $r_{back}$
seems appropriate and we find values of 291 and 242~km~sec$^{-1}$ there from our light curve
solution (which assumed synchronism), in excellent agreement with the line profile results.
Our data support the AG75 values and rule out the smaller values measured by S95.

To confirm the MT method we have applied it to the \ion{He}{1}~5876~\AA\ line of the
B1.5Vn star HIP~77635 that we observed under identical instrumental
conditions soon after TU~Mus on Feb. 22. Its rotational velocity is reported
as 306~km~sec$^{-1}$ in the \cite{bern70} catalogue of
rotational velocities. The HIW of the line turned out to be 8.16~\AA\, which
corresponds to 311~km~sec$^{-1}$ according to Equation \ref{reddening}. This excellent
agreement supports our confidence in the rotational velocities we have
measured for the components of TU~Mus. We have also computed synthetic, broadened
profiles for \ion{He}{1}~4471, 6678 and 7065~\AA\ and the agreement with the observed 
profiles is also excellent.

\subsection{Distance}

Assuming a luminosity ratio $L_2/L_1$=0.55 from our light curve solution, the $V_T$=8.41 magnitude
transforms into Johnson V magnitudes of $V_1$=8.88 and $V_2$=9.52, following \cite{b00}.
With $E_{B-V}$=0.19 (see below) and assuming an O7.5V
classification for the primary, the distance to TU~Mus amounts to 4.77 kpc
using the absolute magnitudes calibrated by \cite{nh94} ($M_V$=--5.1, --4.5 for
O7.5V, O9.5V stars) and a ratio of total to selective absorption of $R_V=3.1$. The same calculation
for the O9.5V secondary results in a distance of 4.86 kpc. We therefore adopt 4.8 kpc
as the distance to TU Mus.

\subsection{Reddening}

\begin{deluxetable}{ccccc}
\tabletypesize{\scriptsize}
\tablecaption{Components of the Interstellar Lines in the TU Mus Spectrum \label{naitable}}
\tablewidth{0pt}
\tablehead{
\colhead{Quantity} & \colhead{Component 1} & \colhead{Component 2} 
           & \colhead{Component 3}  & \colhead{Total}
}
\startdata
RV (km~sec$^{-1}$)   & -15.51$\pm$0.09 &  +5.8$\pm$0.2    & +15.6$\pm$0.1    &                \\               
Width (km~sec$^{-1}$)\tablenotemark{a} & 6.00$\pm$0.04   &5.8$\pm$0.1       &  5.3$\pm$0.2     &                \\
\ion{Na}{1} D2 E.W. (\AA)    & 0.225$\pm$0.002 &  0.216$\pm$0.002 & 0.099$\pm$0.003  &                \\
$E_{(B-V)i}$         & 0.080           &   0.075          &  0.031           & 0.19$\pm$0.01  \\
\enddata
\tablenotetext{a}{Uncorrected for the dominating $\sim$5.0~km~sec$^{-1}$ instrumental width}
\end{deluxetable}

\begin{figure}[!t]
\centerline{\plotone{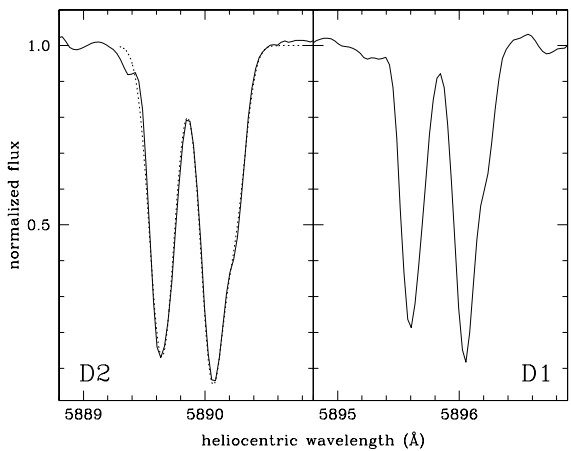}}
\caption{The interstellar lines of \ion{Na}{1}~D (spectrum \#1629) with superimposed the three-component
Gaussian fitting of Table \ref{naitable}. \label{naiplot}}
\end{figure}

The interstellar lines of \ion{Na}{1} offer a means of estimating the
reddening affecting TU Mus independent of the knowledge of the intrinsic
colors. The profile of \ion{Na}{1} D1 and D2 lines on our FEROS spectra of TU~Mus
are shown in Figure \ref{naiplot}. They are clearly composed of three separate
components. Their average position, width and equivalent width have been derived
from multi-Gaussian fitting on all 9 available spectra, and the results are
summarized in Table \ref{naitable}, where the errors are the standard errors. The 9
individual spectra gave extremely consistent results. The multi-Gaussian
{\em mean} fit of Table \ref{naitable} is overplotted on the D2 line of spectrum \#1629 in
Figure \ref{naiplot} to show the accuracy of the fit.

The equivalent widths of the three components have been transformed in
Table \ref{naitable} into the corresponding color excess using the relation calibrated by
\cite{mun97}, with a total color excess amounting to
$E_{B-V}$=0.19$\pm$0.01. Similar profiles are observed for \ion{K}{1} 7698~\AA\ and
H \& K \ion{Ca}{2} interstellar lines. The \ion{Na}{1} widths in Table \ref{naitable} are not corrected
for instrumental width, which amount to 5.0 km~sec$^{-1}$ from unresolved
telluric lines in the vicinity of \ion{Na}{1} lines (thus intrinsic width does not
exceed 3~km~sec$^{-1}$). The line of sight to TU~Mus therefore crosses three
kinematically very sharp and well separated sources of reddening, of pure
interstellar origin. In fact none of the components show radial velocity
change with orbital phase and none share the -4~km~sec$^{-1}$ systemic
velocity.

Of interest is the heliocentric radial velocity of the three components. The
accuracy reported in Table \ref{naitable} is not fictitious because nearby telluric
absorptions show the same wavelength stability from one spectrum to the
other, and it matches the expected high FEROS spectrograph stability
suitable for extra-solar planet searches. The limited resolution of the AG75
spectra did not allow them to recognize the multi-component structure of
the interstellar lines (H and K lines of \ion{Ca}{2}), for
which they estimated a $-7.3\pm0.8$~km~sec$^{-1}$ radial velocity. Averaging
the three components in Table \ref{naitable} weighted according to equivalent
width, a radial velocity of $-0.7\pm0.2$~km~sec$^{-1}$ is 
derived for an unresolved interstellar \ion{Na}{1} profile. However, a
$-7.2\pm0.4$~km~sec$^{-1}$ value averaged by equivalent width is obtained on our spectra
for both H \& K \ion{Ca}{2} as well as \ion{K}{1} interstellar lines. While showing exactly
the same two major components of \ion{Na}{1}~D1 \& D2, they fail to show the third,
reddest one.

\begin{figure}[!b]
\centerline{\plotone{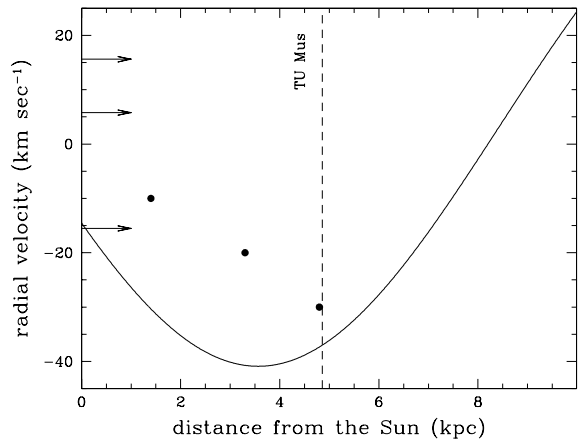}}
\caption{The radial velocity of the three components of \ion{Na}{1} interstellar lines (arrows) compared with
behaviour of the general galactic rotation curve of Hron (1987, continous line) and the local mapping
by Brand and Blitz (1993, solid points) up to the distance of TU~Mus (vertical dashed line).\label{galrot}}
\end{figure}

\noindent

The radial velocities of the three components are plotted in Figure \ref{galrot} where they are compared with
the expected values of the mean galactic rotation along the line of sight to
TU~Mus ($l$=284.8, $b$=--4.1 deg) from the \cite{hron87} formalism with
$A$=--17~km~sec$^{-1}$ and $\alpha$=--2.0~km~sec$^{-1}$~pc$^{-2}$ and
\cite{pont94} values for ($u_\circ$, $v_\circ$, $w_\circ$). It is evident
that the mean Galaxy rotation curve does not reproduce the expected
velocities up to the $\sim$4.8~kpc distance to TU~Mus (note that the line of
sight to the target is still mostly within the $z=\pm 150$~pc dust layer around the
galactic plane at the estimated $\sim$4.8~kpc distance). The discrete
velocity field maps of \cite{brand93}, based on \ion{H}{2} regions and reflection nebulae, do not perform much better, which 
could be anticipated given the scant and widely spaced data used in their
analysis in that part of the Galaxy. Only the stronger
component can be accounted for and located at $\sim$2.5~kpc distance by the \cite{brand93} map. The limited individual reddening implied by the equivalent widths of the three
interstellar lines suggests that only a few limited and 
kinematically scattered bubbles of the interstellar medium are crossed by
the line of sight, with a diffuse component playing a marginal role, if any.
The kinematic information from the interstellar lines cannot
therefore be used to constrain the distance to TU~Mus.

\subsection{Radial velocities}

\begin{figure}
\centerline{\plotone{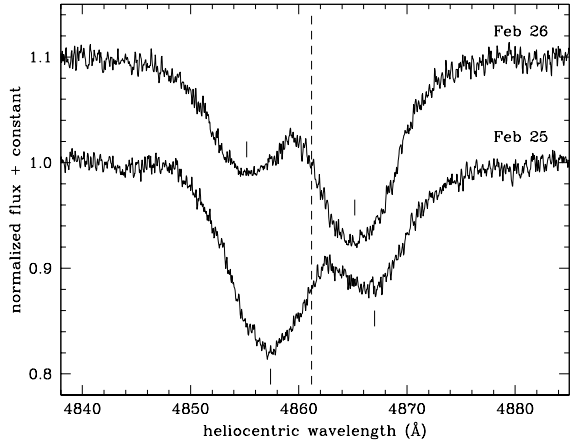}}
\caption{The H$\beta$ profile of TU~Mus for the averaged Feb 25 and Feb 26 spectra of Table~1.
The ticks mark the primary and secondary radial velocity from Table \ref{rvtable}. \label{hbetaplot}} 
\end{figure}

The radial velocity of TU Mus has been measured on the two groups of spectra
in Table \ref{speclist} obtained close to quadrature phases. Those around orbital phase 0.03 are very close to conjunction and thus have insufficient 
velocity separation between the two components to obtain reliable radial 
velocities.
Line splitting at quadrature is wide and allows an easy and firm determination 
of the radial velocity, as shown for H$\beta$ in Figure \ref{hbetaplot}. The radial velocities
have been obtained by Gaussian de-convolution of the profiles into two
components, which provide an excellent overall fit to the observed line
profile. The measured lines are \ion{He}{1}~4016.218, 4471.507, 4921.929, 5015.675,
5875.651, 6678.149, 7065.276~\AA; \ion{He}{2}~4199.831, 4541.589, 4685.682,
5411.524~\AA; and H$\gamma$, H$\beta$, H$\alpha$. The mean
of the radial velocities (and associated standard error) derived from these
14 lines is in Table \ref{rvtable}. 

\begin{deluxetable}{cccccccc}
\tablecaption{Heliocentric radial velocities of TU~Mus \label{rvtable}}
\tablewidth{0pt}
\tablehead{
   \colhead{Spectrum Number} & \colhead{Orbital Phase} & \multicolumn{3}{c}{Primary} 
        & \multicolumn{3}{c}{Secondary}\\
   \multicolumn{2}{c}{} & \colhead{This Paper} & \colhead{AG75} & \colhead{S95} 
        & \colhead{This Paper} & \colhead{AG75} & \colhead{S95}\\
}
\startdata
1629-30-31 & 0.211 &$-244\pm5$  &$-246$ & $-225$ & $+350\pm5$ & $+364$ & $+324$ \\ 
1693-94-95 & 0.771 &$+237\pm4$  &$+249$ & $+197$ & $-380\pm8$ & $-362$ & $-348$ \\ 
\enddata
\end{deluxetable}

\begin{figure}
\centerline{\plotone{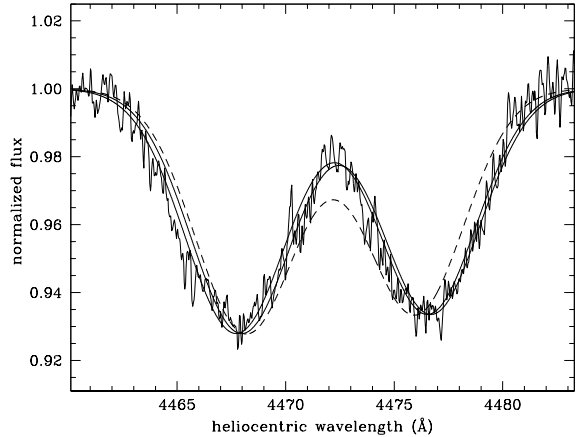}}
\caption{The HeI~4471~\AA\ profile for Feb 25 spectra of Table \ref{speclist}, superposed with
the double Gaussian fitting with our velocities from Table \ref{rvtable} (thick solid line), corresponding
AG75 velocities (thin solid line) and S95 ones (dashed line). \label{heiplot}}
\end{figure}

Our high dispersion, high S/N radial velocities are very close to the photographic ones by AG75 and definitively rule out the smaller 
ones obtained by S95 on low S/N IUE SWP high resolution
spectra. This is clearly demonstrated by Figure \ref{heiplot} that shows the \ion{He}{1}~5876~\AA\ line of spectrum \# 1629
superimposed with the two component fit using our radial velocities in Table \ref{rvtable} 
and the corresponding values for AG75 and S95. 

\subsection{Spectral classification and metallicity}

Both stars are O-type, showing the whole series of expected \ion{He}{2} lines. Their
spectral types are somewhat different, as the comparison of H$\beta$ and
\ion{He}{1}~4471~\AA\ profiles in Figures \ref{hbetaplot} and \ref{heiplot} suggests. The ratio of the two
H$\beta$ lines is 1.8 while those of the \ion{He}{1} is 1.3, favoring a two
spectral subtype higher temperature for the primary. A reasonable estimate seems to
be O7.5 for the primary and O9.5 for the secondary.

Other than the hydrogen and helium lines, only two \ion{C}{3} lines are clearly 
visible at 4650~\AA\ and 4069~\AA\. (\ion{Si}{4} lines are unsuitable in this regard.) Both \ion{C}{3}
lines are actually blends of three individual lines with complicated
profiles caused by wide wavelength separation within the triplets, fast
rotation, and the SB2 nature of the spectra. However, a comparison with synthetic
spectra does not support a carbon content, and therefore metallicity,
far from the solar value.

\section{Photometry}

TU Mus was observed by RHN on 10 nights from 2002 April 13 to June 4 at the 
Mount John University Observatory while he was a guest of the 
University of Canterbury in Christchurch, New Zealand. The f/13.5 Cassegrain focus on the 0.61 meter Optical Craftsmen telescope was used together 
with a Santa Barbara Instrument Group ST-9E CCD camera, utilizing the KAF-0261E chip (512x512) 
with 20 micron pixels.  A telecompressor lens operated at 2.11x compression reduced the focal 
ratio to 6.4 and gave a field of view of 9' x 9'.  The B and V (Johnson) filters were by Schuler. 

Reduction of the CCD images and aperture photometry was done with the MIRA software package.\footnote{Available at http://www.axres.com}  
Bias and dark removal plus flat fielding (dome flats) were done in the usual way.   The comparison 
star was GSC 8984:0579 and the check stars were GSC 8984:1473 and GSC 8984:1623. Weather
and time constraints did not permit the determination of transformations to 
the standard system. There were 1973 frames taken in B and 1962 in V.  However, wispy clouds often caused scattered 
measurements.  In order to remove errant measurements in an objective way, the raw comparison magnitudes 
were plotted versus time.  Points  adjacent to rapidly changing values (greather than 0.2 magnitudes per 
sampling interval of 30 seconds) were rejected.  In this way, the data set was reduced to 1148 
points in B and 1493 points in V, and the light curves were cleaned up considerably.

\section{Data Analysis}
\label{analysis}

We performed a simultaneous analysis of the AG75 photometry and radial velocities, 
the Hipparcos photometry, and our new photometry and velocities with the 2003 version of the 
Wilson-Devinney (WD) code. This version of WD has the capability of modeling the radiation of 
the stars by \cite{k93} atmospheres and we used this feature in our solutions. 

\begin{figure}
\centerline{\plotone{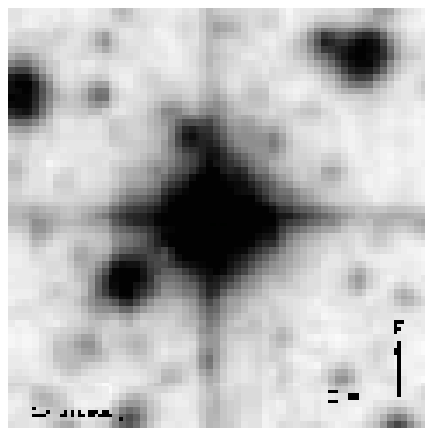}}
\caption{The Digitized Sky Survey image of the TU Mus field showing the companion approximately
$15"$ to the southeast. \label{companionplot}}  
\end{figure}

In early exploratory fits with
the light curve program of WD, it became clear that the AG75 photometry, the Hipparcos
photometry, and our new photometry were not consistent in terms of the depths of the eclipses. The 
AG75 and Hipparcos photometry matched well, but our new photometry had slightly deeper eclipses,
requiring an inclination about one degree higher. Rather than indicating some real change in the 
inclination of the system, the differences result from the fact that the older photometry suffers 
from third light contamination.
The Digitized Sky Survey image of the field shows a companion about $15"$ southeast of TU Mus. Our
aperture photometry was done with a $10"$ aperture and therefore did not include the companion. The 
effective aperture for the Hipparcos observations was about $30"$ \citep{hipp}, thus including the 
companion. The AG75 data also included the companion since their aperture was $30"$.
The Guide Star Catalog lists the brightness of the companion as $13^{m}_{.}0 \pm 0^{m}_{.}4$. 
In our light curve solutions below, we allowed for third light in all of the photometry but
the values for the $BV$ curves were always very tiny, thus we conclude that third light does not
affect our $BV$ data. The solution did find third light in the AG75 and Hipparcos data and the results
are consistent with the GSC brightness of the companion.  Figure \ref{companionplot}
shows the DSS image of the area around TU Mus.

Previously, WD had been
modified to allow for the use of either binary phase or time as the independent variable 
\citep{wt98}. In the latter case, one can solve for the orbital period ($P$), its first time derivative 
($\dot{P}$), and the reference epoch ($HJD_0$, usually the time of primary minimum). In order to investigate 
the behavior of the period of TU~Mus, we used time as the independent variable for our simultaneous
solutions, adjusting the period 
and reference epoch. Attempts to adjust \.{P} always resulted in values statistically 
indistinguishable from zero, indicating that the period of TU~Mus has remained constant over
the thirty year span of the observations being analyzed. Our estimate of the ephemeris is
\begin{equation}
\label{ephem}
HJD_{min} = 2448500.3080(1) + 1.38728653(2) \times E
\end{equation}
for primary eclipse where the quantities in parentheses are the $1\sigma$ uncertainties in the last digits of 
the parameters. 

Other parameters adjusted in the simultaneous solution were the semi-major axis of the relative orbit
($a$), the binary center of mass radial velocity ($V_{\gamma}$), orbital inclination ($i$),
secondary mean effective temperature ($T_2$), common envelope surface potential ($\Omega$), mass ratio
($q$), and the bandpass-specific luminosity of the primary ($L_1$). Other parameters, such as the
bolometric albedos and gravity brightening exponents, were held fixed at their expected theoretical 
values. The square root limb darkening law was used with coefficients from \cite{wvh93}. The 
mean effective temperature of the primary was set to 35,000 K based on the O7.5 spectral type. Data set weights were determined by the scatter of the observations. WD gives output at each step
from which the standard deviations of the data sets, $\sigma$, can be calculated. It is sufficient to determine reasonably 
good starting values for the standard deviations by examining plots of the
observations and then adjust them as the solution progresses. Within a given light curve
or radial velocity curve, WD can also apply individual weights. For the radial velocities,
we set the base curve weights for the AG75 velocities and then gave our new velocities higher individual 
weights based on their estimated errors. In order to eliminate potential local minimum problems, we 
began the solutions from several starting points and found that the same minimum was always 
recovered. 

\begin{deluxetable}{cc}
\tabletypesize{\scriptsize}
\tablecaption{Parameters of TU Mus \label{parameters}}
\tablewidth{0pt}
\tablehead{
\colhead{Parameter} & \colhead{Value\tablenotemark{a}}
}
\startdata
$a$ & $17.7 \pm 0.2 R_{\sun}$  \\
$V_{\gamma}$ & $-4 \pm 4$ km sec$^{-1}$\\
$i$ & $77^{\circ}_{.}8 \pm 0^{\circ}_{.}1$  \\
$T_{1}$ & $35,000$ K \\
$T_{2}$ & $31,366 \pm 16$ K\\
$\Omega_{1}$ & $3.137 \pm 0.002$ \\
$q$ &  $0.651 \pm 0.001$ \\
$HJD_{0}$ &  $2448500.3066 \pm 0.0001$ \\
$P$ & $1.38728653 \pm 0.00000002$ days \\
$L_{1}/(L_{1}+L_{2})_{H_P}$ & $ 0.648 \pm 0.006$  \\
$L_{1}/(L_{1}+L_{2})_u$ & $ 0.659 \pm 0.003$  \\
$L_{1}/(L_{1}+L_{2})_v$ & $ 0.648 \pm 0.003$  \\
$L_{1}/(L_{1}+L_{2})_b$ & $ 0.646 \pm 0.003$  \\
$L_{1}/(L_{1}+L_{2})_y$ & $ 0.646 \pm 0.003$  \\
$L_{1}/(L_{1}+L_{2})_B$ & $ 0.648 \pm 0.003$  \\
$L_{1}/(L_{1}+L_{2})_V$ & $ 0.646 \pm 0.003$  \\
$l_{3} (H_P)$ & $ 0.030 \pm 0.007$ \tablenotemark{b} \\
$l_{3} (u)$ & $ 0.023 \pm 0.004$  \\
$l_{3} (v)$ & $ 0.023 \pm 0.004$  \\
$l_{3} (b)$ & $ 0.025 \pm 0.004$  \\
$l_{3} (y)$ & $ 0.026 \pm 0.004$  \\
$R_{1}$ & $7.48 \pm 0.08$ R$_{\sun}$ \\
$R_{2}$ & $6.15 \pm 0.07$ R$_{\sun}$ \\
$M_{1}$ & $23.5 \pm 0.8$ M$_{\sun}$ \\
$M_{2}$ & $15.3 \pm 0.4$ M$_{\sun}$ \\
log $L_{1}/L_{\sun}$ & $4.8 \pm 0.2$ \tablenotemark{c} \\
log $L_{2}/L_{\sun}$ & $4.5 \pm 0.2$ \\
\enddata
\tablenotetext{a}{Quoted errors are 1-$\sigma$ errors.}
\tablenotetext{b}{$l_3$ values are in units of total system light at phase 0.25.}
\tablenotetext{c}{The luminosity errors are estimates based on the uncertainty of a few thousand K
in the effective temperature of the primary. The error contributions to the luminosities due to the
errors in the radii are more than an order of magnitude smaller.}
\end{deluxetable}

\begin{figure}
\centerline{\plotone{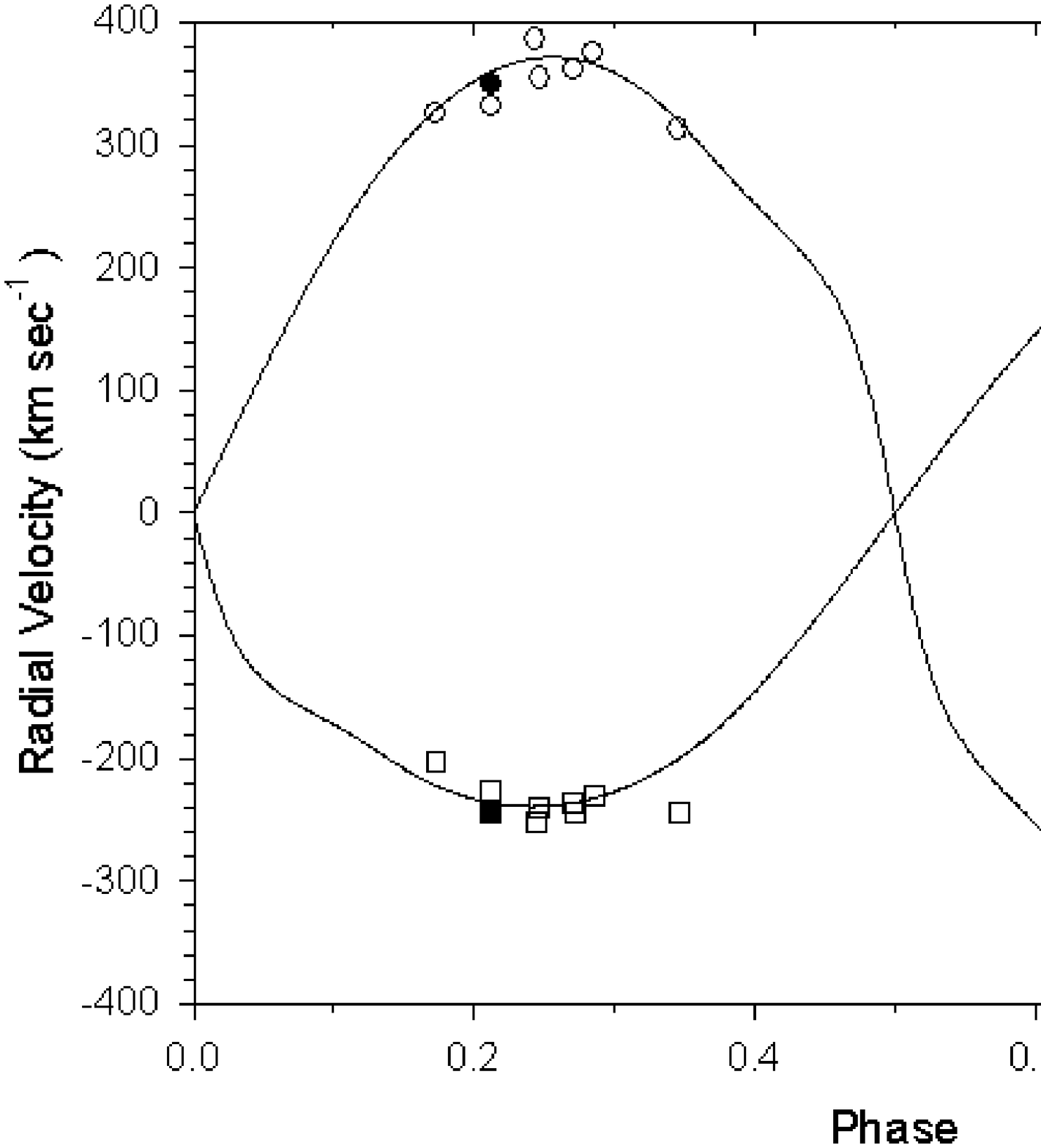}}
\caption{The fit to the radial velocities from this paper (solid symbols) and those of AG75 (open
         symbols). \label{rvplot}}  
\end{figure}

\begin{figure}
\centerline{\plotone{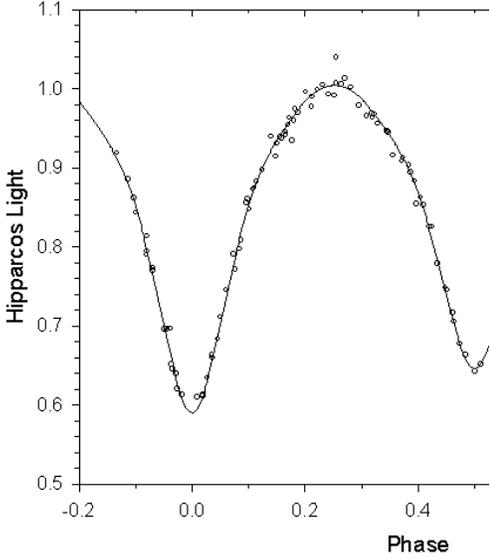}}
\caption{The fit to the Hipparcos observations using the parameters of Table \ref{parameters}. 
    \label{hippplot}} 
\end{figure}

\begin{figure}
\centerline{\plotone{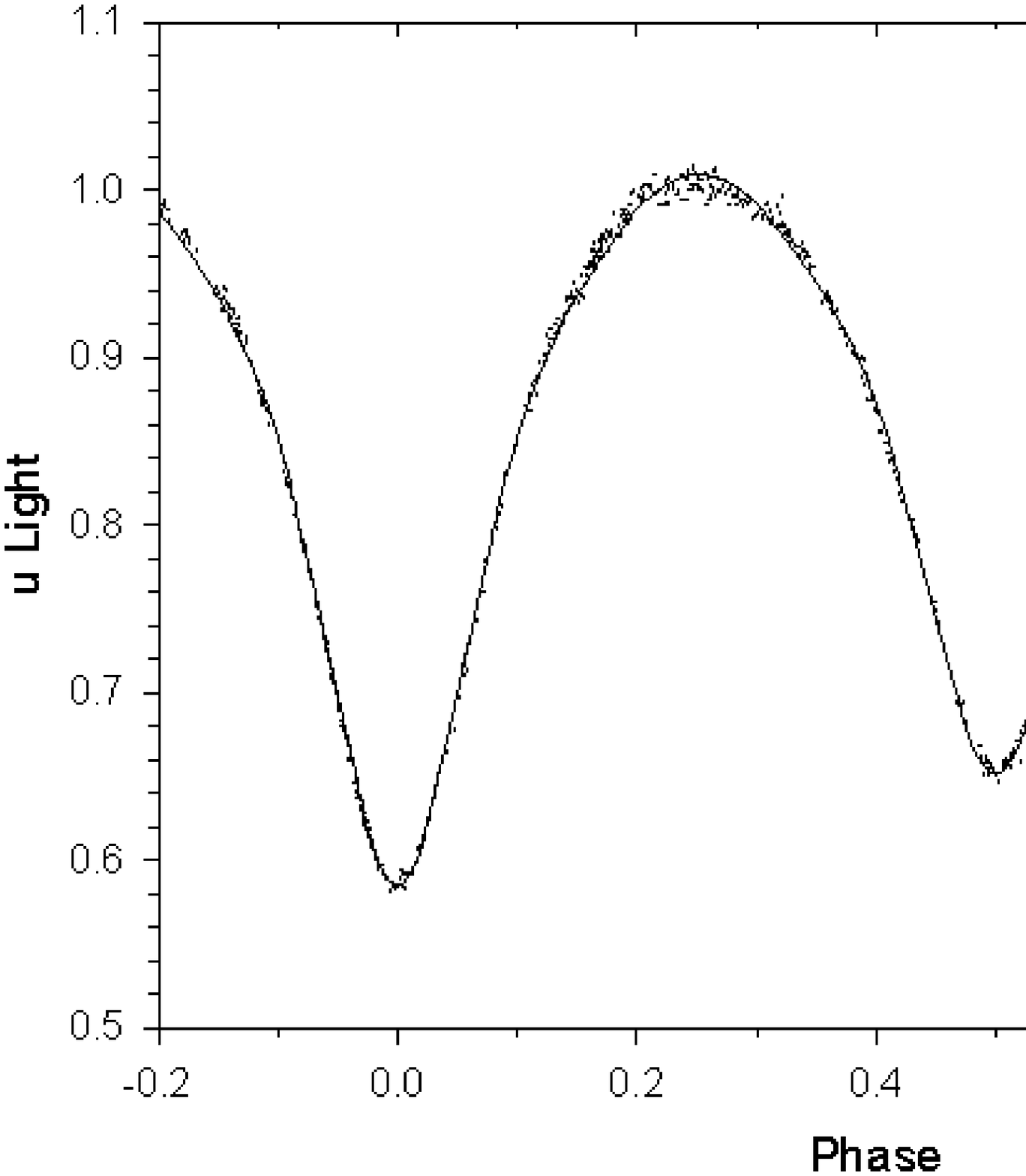}}
\caption{The fit to the AG75 $u$ observations using the parameters of Table \ref{parameters}. 
   \label{uplot}} 
\end{figure}

\begin{figure}
\centerline{\plotone{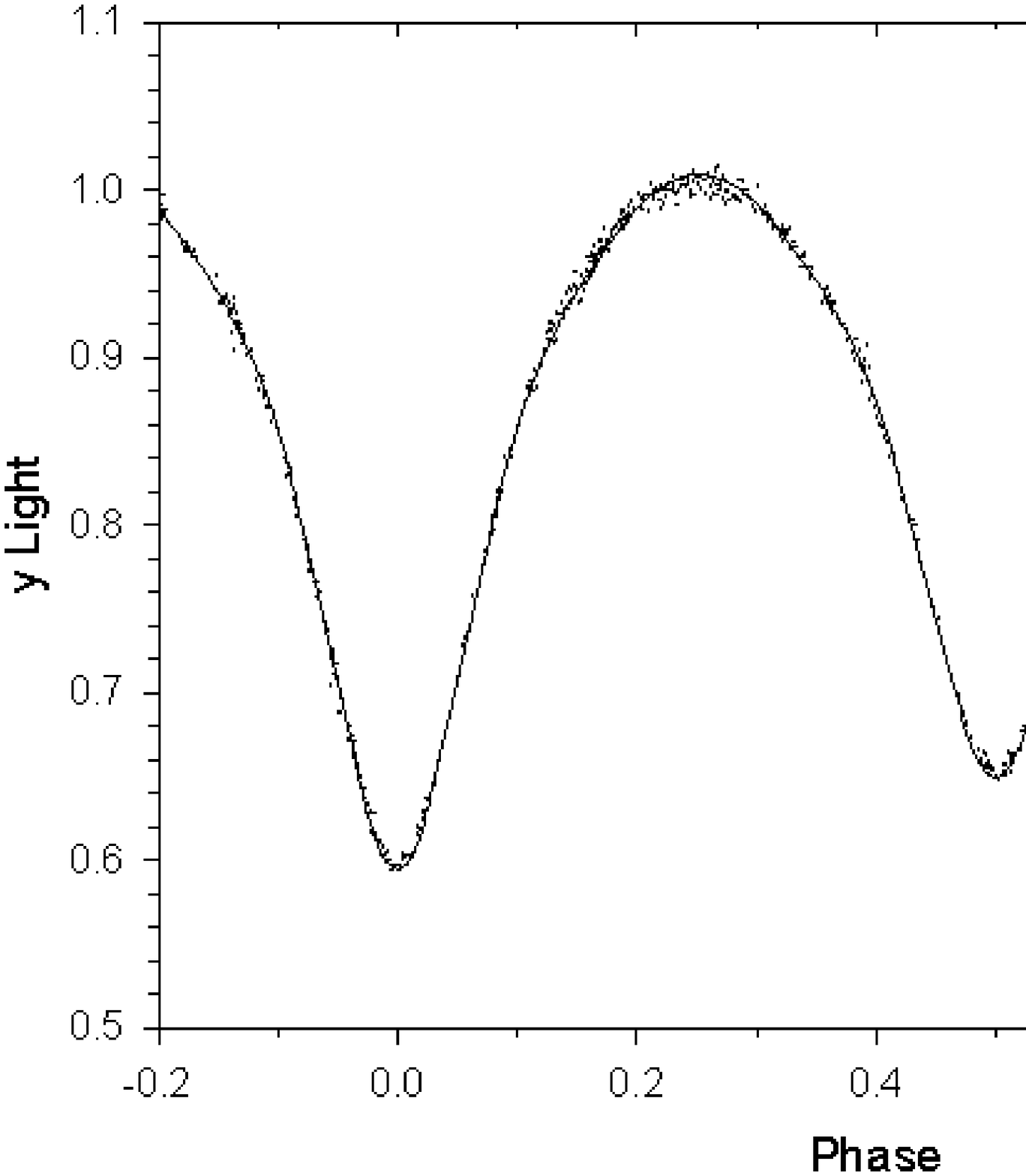}}
\caption{The fit to the AG75 $y$ observations using the parameters of Table \ref{parameters}. 
    \label{yplot}} 
\end{figure}

\begin{figure}
\centerline{\plotone{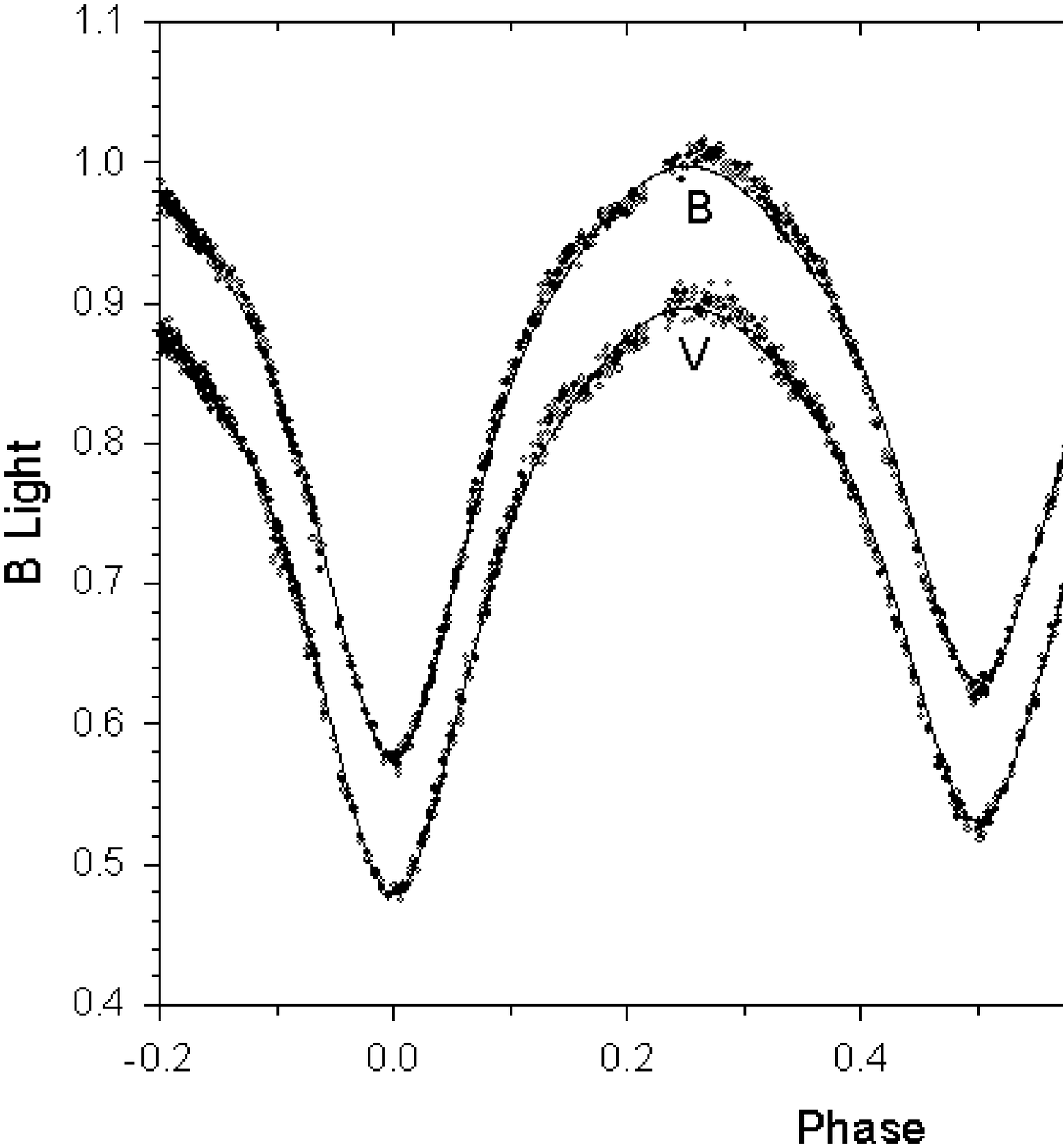}}
\caption{The fit to our new $BV$ observations using the parameters of Table \ref{parameters}. 
    \label{rhnplot}} 
\end{figure}

Table \ref{parameters} shows the results of the simultaneous solution. TU~Mus is just barely in an 
overcontact configuration, having a surface potential of 3.137 $\pm$ 0.002 while the inner critical potential 
is 3.156 for the estimated mass ratio of 0.651. Using the degree of contact index defined by \cite{wr81}, 
where point contact has a value of 1.0, we find a value of 1.006 for TU~Mus. Figure \ref{rvplot} shows 
the fit to the AG75 radial velocities and the ones we obtained. The fit to the Hipparcos photometry is 
shown in Figure \ref{hippplot} and since all of the various observations were phased with the 
ephemeris given in Equation \ref{ephem}, it shows that the period has remained constant to the precision with which
we can measure it.  Figure \ref{uplot} shows the fit to the AG75 $u$ observations and exhibits a small 
but noticeable asymmetry in the maxima, a phenomenon frequently seen in the light curves of
late-type overcontact (W UMa) systems. As can be seen in Figure \ref{yplot}, the asymmetry is
slightly smaller in the $y$ data, indicating that the source of the asymmetry is a high temperature phenomenon,
perhaps a superluminous area as suggested in \objectname{CE Leo} by \cite{sam93} and \objectname{YZ Phe}
by \cite{st95}. Figure \ref{rhnplot} shows the fits to our $B$ and $V$ data. An asymmetry is present in 
these newer data but in the opposite sense from the AG75 data, indicating that the source of the 
asymmetry is variable in time.

\section{Age and Evolutionary Status}

Very little is known about the structure and evolution of overcontact binaries with radiative 
envelopes. To our knowledge, no models of this type have been published. Although obviously
limited, a comparison with single-star models can be instructive in estimating the age
and evolutionary status of TU Mus. Using a stellar evolution code (\cite{han94}, 
\cite{ppe73}, \cite{ppe72}, \cite{ppe71}) kindly supplied to us by P. Eggleton, we have
constructed evolutionary models of $23.5 M_{\sun}$ and $15.3 M_{\sun}$. The model for the 
primary reaches the observed radius of the TU Mus primary at about 2.6 Myr and the model for the 
secondary reaches the observed radius at about 6.2 Myr. Because of the luminosity exchange 
from the primary to the secondary, the radius of the primary is probably less than it would be 
without the luminosity transfer. Similarly, the secondary is probably larger than it otherwise
would be. Thus the true age of TU Mus probably lies between these two values. Our ignorance
of the structure and evolution of these stars is so great, however, that one cannot discount
the possibility that the current configuration is a result of a much more complicated 
evolutionary history.

Two results from this work, the overcontact 
configuration and the lack of any period change, have interesting implications
for the structure of early-type overcontact systems. In the late-type
overcontact (W UMa) systems, the luminosity transfer drives mass transfer from
the less massive to the more massive star, resulting in period changes. But
if TU Mus truly is in an overcontact configuration, the same level of
mass transfer cannot be taking place because of the constant period. What is clearly needed is more work in this area 
with modern codes like Djehuty \citep{ppe03}. More sophisticated
modeling by 3D stellar evolution codes will certainly give us a better understanding of the
structure and evolution of early-type overcontact systems and the well-determined properties
of TU Mus makes it a good target of such modeling. 

\section{Conclusions}

Our analysis of new and existing data on TU Mus shows it to be an overcontact binary consisting
of a $23.5 M_{\sun}$ primary and a $15.3 M_{\sun}$ secondary with a small degree of contact.
Our results are in very good agreement with the results of AG75 and contradict the 
smaller masses found by S95. The source of the discrepancy between the UV and optical velocities
is not clear. One idea we explored was the effect of the change in the center of light due to
the reflection effect between the UV and optical, but our experiments indicated that this effect
is much too small to explain the large discrepancy. Given our agreement with the AG75 velocities,
one is left with two possible conclusions: that the UV velocities suffer from some unrecognized
systematic error or there is some fundamental effect that causes a difference between UV and 
optical velocities for stars of this type. We note that Stickland and colleagues have seen a 
similar discrepancy for LY Aur \citep{s94}.
Our current understanding of the structure and evolution of 
early-type overcontact binaries is quite limited and we hope that our results for the observed 
properties of TU Mus may serve as a good test of future 3D models of these stars.

\acknowledgments
RHN thanks the American Association of Variable Star Observers for the loan
of the ST-9E camera. RHN would also
like to thank specifically the two technicians at the observatory, Alan
Gilmore and Pam Kilmartin, for their superb help, and the members of the
Physics and Astronomy Department at the University of Canterbury for
their warm hospitality. This research has made use of the SIMBAD database, operated at CDS, 
Strasbourg, France.

\clearpage


\end{document}